\documentclass[11pt]{article}
\usepackage{graphicx}
\usepackage{cite}

\newcommand{ \be}{\begin{equation}}
\newcommand{ \ee}{\end{equation}}
\newcommand{ \bea}{\begin{eqnarray}}
\newcommand{ \eea}{\end{eqnarray}}

\begin{document}

\begin{titlepage}

\begin{center}
\vspace{2cm} {\LARGE\bf Energy Dependence of CP-Violation Reach}

\vspace{5mm} {\LARGE\bf for Monochromatic Neutrino Beam}

\vspace{15mm}
{\large\bf      Jos\'e Bernab\'eu and Catalina Espinoza}

\vspace{13mm}

{\it    IFIC, Universitat de Val\`encia--CSIC, E-46100, Burjassot, Val\`encia, 
	Spain \\[3mm] 
}
        
\vspace{5cm}
{\large\bf Abstract} 
\end{center}

\vspace{7mm} 
\noindent 
The ultimate goal for future neutrino facilities is the determination of CP violation in neutrino oscillations. Besides $\vert U(e3) \vert \ne 0$, this will require precision experiments with a very intense neutrino source and energy control. With this objective in mind, the creation of monochromatic neutrino beams from the electron capture decay of boosted ions by the SPS of CERN has been proposed. We discuss the capabilities of such a facility as a function of the energy of the boost and the baseline for the detector.   We compare the physics potential for two different configurations: I) $\gamma=90$ and $\gamma=195$ (maximum achievable at present SPS) to Frejus; II) $\gamma=195$ and $\gamma=440$ (maximum achievable at upgraded SPS) to Canfranc. We conclude that the SPS upgrade to 1000 GeV is important  to reach a better sensitivity to CP violation iff it is accompanied by a longer baseline.  In both Setups, the gain in the CP violation sensitivity with a previous knowledge of  $\vert U(e3) \vert$ is apparent.

\end{titlepage}


\section{Introduction}

\noindent Neutrinos do have masses and mixings. Present evidence \cite{fukuda, ahmad} from neutrino osci\-lla\-tions is consistently interpreted in terms of two independent mass-differences and mixings: the so-called atmospheric sector (2,3) and the solar sector (1,2). The initial discovery of the zenith effect with atmospheric neutrinos led to the (2,3) sector in neutrino oscillations, later confirmed by long baseline experiments with accelerator neutrinos. The solar neutrino solution to the historical solar neutrino problem led to the (1,2) sector from neutrino oscillations in solar matter, later confirmed by long-baseline vacuum oscillations with reactor neutrinos. Whereas the sign of $\Delta m_{12}^2$ is thus known, the determination of the sign of $\Delta m_{23}^2$ needs the incorporation of matter effects and then $\vert U(e3) \vert \ne 0$. The two mixings angles are large: $\theta_{23}$ could even be $45^{\circ}$, whereas $\theta_{12}$ is large although not maximal.

The third connecting mixing $\vert U(e3) \vert$ is bounded as  $\theta_{13}\le 10^{\circ}$   from the CHOOZ
reactor experiment \cite{apollonio}. 
The angle $\theta_{13}$ remains  thus undetermined. The approved reactor experiments Double CHOOZ \cite{Lasserre:2004vt} and Daya-Bay \cite{Guo:2007ug}, as well as the second generation of long-baseline superbeam experiments T2K \cite{t2k} and NOVA \cite{Ayres:2004js} will address this point. A number of experimental facilities to significantly improve on present
sensitivity and look for CP-violation have been discussed in the literature: neutrino factories (neutrino beams from boosted-muon 
decays)~\cite{geer,drgh,nufact}, 
superbeams (very intense conventional neutrino beams)~\cite{JHF,NUMI,splcern,superbeam},
improved reactor experiments~\cite{reactor_deg} 
and  $\beta$-beams \cite{zucchelli}.
The original standard scenario for beta beams with low $\gamma=60/100$ and  short baseline $L=130$~Km from CERN to Frejus with $^6He$ and $^{18}Ne$ ions can have a variant by using an electron capture facility for monochromatic neutrino beams \cite{Bernabeu:2005jh}. New proposals also include the high $Q$ value $^8Li$ and $^8Be$ isotopes in a $\gamma=100$ facility \cite{Rubbia:2006pi}. For the standard beta beam facility, a study of the physics reach as function of the boost and the baseline has been made \cite{Burguet-Castell:2005pa}. In this paper we discuss the physics reach that a high energy facility for  EC beams may provide with the expected SPS upgrade at CERN.
In Section~2 we discuss the virtues of the suppressed oscillation channel $(\nu_e \to \nu_\mu)$ in order to have access to the parameters $\theta_{13}$ and $\delta$. To disentangle the CP-violating phase we emphasize the method of using energy dependence, as obtainable in the EC facility.  In Section~3 we present new results on the comparison between (low energies, short baseline) and (high energies, long baseline) configurations for an EC facility with a single ion. Section~4 gives our conclusions.

\vspace{0.5cm}

\section{CP-even and CP-odd terms}

\noindent The observation of $CP$ violation needs an experiment in which the emergence of another neutrino flavour
 is detected rather than the deficiency of the original flavour of the neutrinos. At the same time, the interference needed to generate CP-violating observables can be enhanced if both the atmospheric and solar components have a similar magnitude. This can happen in the suppressed $\nu_e \to \nu_{\mu}$ transition. The appearance
 probability $P(\nu_e \to \nu_{\mu})$ as a function of the distance between source and detector $(L)$ is given by \cite{cervera}
\begin{eqnarray}\label{prob}
P({\nu_e \rightarrow \nu_\mu})  \simeq ~ 
s_{23}^2 \, \sin^2 2 \theta_{13} \, \sin^2 \left ( \frac{\Delta m^2_{13} \, L}{4E} \right ) \nonumber \\
  + ~   c_{23}^2 \, \sin^2 2 \theta_{12} \, \sin^2 \left( \frac{ \Delta m^2_{12} \, L}{4E} \right ) 
\nonumber \\
 + ~ \tilde J \, \cos \left ( \delta - \frac{ \Delta m^2_{13} \, L}{4E} \right ) \;
\frac{ \Delta m^2_{12} \, L}{4E} \sin \left ( \frac{  \Delta m^2_{13} \, L}{4E} \right ) \,,
\end{eqnarray}
where $\tilde J \equiv c_{13} \, \sin 2 \theta_{12} \sin 2 \theta_{23} \sin 2 \theta_{13}$.
   The three terms of Eq.~(\ref{prob}) correspond, res\-pec\-ti\-ve\-ly, to contributions from the
atmospheric and solar sectors and their interference. As seen, the $CP$ violating contribution
has to include all mixings and neutrino mass differences to become observable. The four measured parameters $(\Delta m_{12}^2,\theta_{12})$  and  $(\Delta m_{23}^2,\theta_{23})$ have been fixed throughout this paper to their mean values \cite{Gonzalez-Garcia:2004jd}. 

\begin{figure}[ht]
\centering
\includegraphics[width=9.5cm, angle=-90]{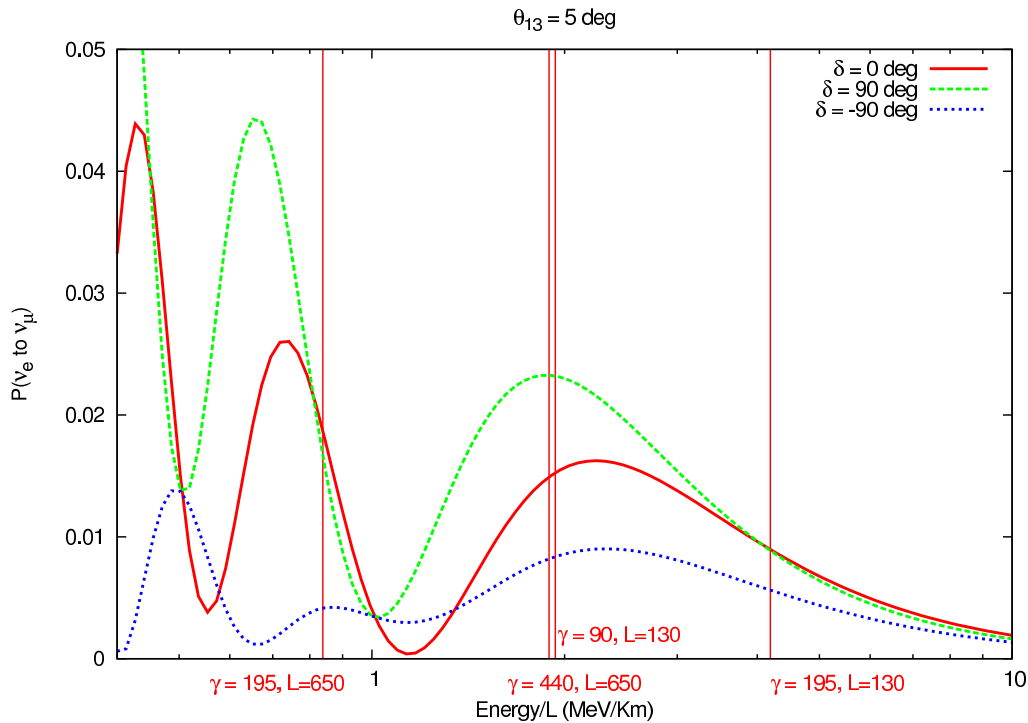}
\caption{\label{proba}The appearance probability $P(\nu_e \rightarrow \nu_{\mu})$ for neutrino oscillations
 as a function of the LAB energy E/L, with fixed  connecting
mixing. The three curves refer to different values of the CP violating phase $\delta$. The vertical lines are the energies of our simulation study in the EC facility.}
\end{figure}

Neutrino oscillation phenomena are energy dependent (see Fig.\ref{proba}) for a fixed distance between source and detector, and the observation of this energy dependence would disentangle the two important parameters: whereas $\vert U(e3) \vert$ gives the strength of the appearance probability, the $CP$ phase $\delta$ acts as a phase-shift in the interference pattern. In fact, a general theorem \cite{Cabibbo}  states that, under the assumptions of CPT invariance and absence of absorptive parts, the CP-odd probability is odd in time or, equivalently, odd in $L$. As vacuum oscillations depend on $L/E$, this result implies that, for fixed $L$,  the CP-odd probability is odd in $E$, whereas the  CP-even terms are even in $E$. This is satisfied by Eq.~(\ref{proba}), with three contributions, the atmospheric, the solar and their CP-even interference, which are even functions of $E$, and the CP-odd interference, which is odd in $E$. These properties suggest the consideration of a facility able to study the detailed energy dependence by means of fine tuning of a boosted monochromatic neutrino beam. In an electron capture facility the neutrino energy is dictated by the chosen boost
of the ion source and the neutrino beam luminosity is concentrated at a single
known energy which may be chosen at
will for the values in which the sensitivity for the $(\theta_{13}, \delta)$ parameters is higher. This is in contrast to beams with a continuous spectrum, where the intensity is shared between sensitive
and non sensitive regions. Furthermore, the known definite energy would help in the control of both the systematics
and the detector background. In the beams with a continuous spectrum, the neutrino energy has to be reconstructed in the detector. In water-Cerenkov detectors, this reconstruction is made from supposed quasielastic events by measuring both the energy and direction of the charged lepton. This procedure suffers from non-quasielastic background, from kinematic deviations due to the nuclear Fermi momentum and from dynamical suppression due to exclusion effects \cite{Bernabeu72}.

The above discussion proves that the study of neutrino oscillations in terms of neutrino energy will be able to separate out the CP phase $\delta$ from the mixing parameters. A control of this energy may be obtained from the choice of the boost in the EC facility with a single ion. In order for this concept to become operational, it is necessary to combine it with the recent discovery of nuclei  far from the stability line, having super
 allowed spin-isospin transitions to a giant Gamow-Teller resonance kinematically accessible
 \cite{algora}. Thus the rare-earth nuclei above $^{146}Gd$ have a small enough half-life to allow 
 electron capture processes in the decay ring. This is in contrast with the proposal of EC beams  with fully stripped long-lived ions \cite{Sato:2005ma}.  We discuss the option of short-lived ions \cite{Bernabeu:2005jh}.

\vspace{0.5cm}

\section{Physics reach at different energies and baselines}

\noindent Electron Capture is the process in which an atomic electron is captured by a proton
of the nucleus leading to a nuclear state of the same mass number $A$, replacing the 
proton by a neutron, and a neutrino. Its probability amplitude is proportional to the atomic 
wavefunction at the origin, so that it becomes competitive with the nuclear $\beta^+$ decay 
at high atomic number $Z$. Kinematically, it is a two body decay of the atomic ion into a nucleus and the 
neutrino, so that the neutrino energy is well defined and given by the difference between
the initial and final atomic masses minus the excitation energy of the 
final nuclear state. In general, the high  $Z$ nuclear beta-plus decay $(\beta^+)$
 and electron-capture $(EC)$ transitions are very "forbidden", i.e., disfavoured, because the
energetic window open in these channels does not contain the important Gamow-Teller strength
excitation seen in (p,n) reactions. There are a few cases, however, where the Gamow-Teller 
resonance can be populated having the occasion of a strong ``allowed'' transition. For the rare-earth nuclei above $^{146}Gd$, the filling of the intruder level $h_{11/2}$ 
for protons opens the possibility of a spin-isospin transition to the allowed level $h_{9/2}$
for neutrons, leading to a fast decay. Our studies for neutrino beam capabilities have used the $^{150}Dy$ ion with half life of 7.2 min, a Branching Ratio to neutrino channels of $64\%$ (fully by EC) and neutrino energy of $1.4$~MeV in the C.M. frame, as obtained from its decay to the single giant Gamow-Teller resonance in the daugther $^{150}Tb^*$. 

 The parent radioactive ion is accelerated and then accumulated and storaged. A neutrino of energy $E_0$ in C. M. that emerges from the decay in these conditions will
 be boosted in energy. This LAB energy is a function
 of the angle $(\theta)$ of neutrino detection and Lorentz gamma $(\gamma)$ of the ion at the moment of decay and it can 
be expressed as $E = E_0 / [\gamma(1- \beta \cos{\theta})]$. The angle $\theta$ here expresses the deviation between the actual neutrino detection and the ideal detector
 position in the prolongation of one of the long straight sections of the decay ring.
 The neutrinos emerging from a boosted ion beam decaying by EC are concentrated inside a narrow cone around the forward direction. If the ions
 are kept in the decay ring longer than the half-life, the energy distribution of the Neutrino 
Flux arriving to the detector in the forward direction, in absence of neutrino oscillations, is given by the Master Formula

\begin{eqnarray}\label{master}
\frac{d^2N_\nu}{ dS dE}
& = & \frac{1}{\Gamma} \frac{d^2\Gamma_\nu}{dS dE} N_{ions} \nonumber \\
& \simeq & \frac{\Gamma_\nu}{\Gamma} \frac{ N_{ions}}{\pi L^2} \gamma^2
\delta{\left(E - 2 \gamma E_0 \right)},
\end{eqnarray}
with a dilation factor $\gamma >> 1$. It is remarkable that the result is given only in terms of the branching ratio for electron capture
and the neutrino energy and independent of nuclear models. In Eq.~(\ref{master}), $N_{ions}$ is the total number of 
 ions decaying to neutrinos. At the first oscillation maximum, with $E / L$ fixed, Eq.~(\ref{master}) says that lower neutrino energies $E_0$ in the
 proper frame give higher neutrino fluxes.  The number of events will increase with higher
 neutrino energies as the cross section increases with energy. To conclude, in the forward 
direction the neutrino energy is fixed by the boost $E = 2 \gamma E_0$, with the entire neutrino 
flux concentrated at this energy.
 As a result, such a facility will measure the neutrino 
oscillation parameters as a function of energy  by changing the $\gamma$'s of the decay ring (energy dependent measurement)
 and there is no need of energy reconstruction in the detector. In this situation, the experiment becomes a counting-rate of events.
\begin{figure}[htb]
\centering
\includegraphics[width=6.3cm, angle=-90]{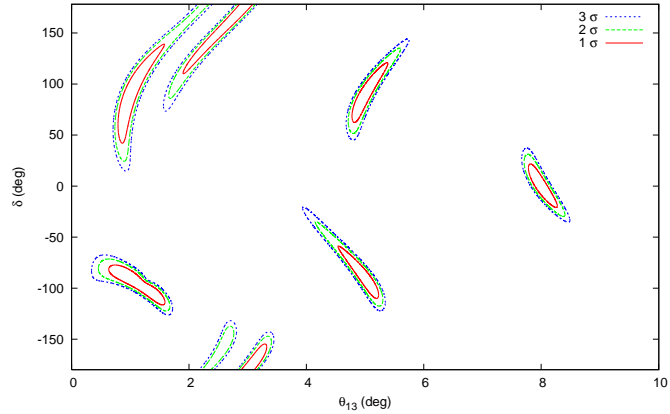}
\caption{\label{fit-setupI}Setup I as explained in the text. Fit for $(\theta_{13}, \delta)$ from statistical distribution of events with assumed values of the parameters.}
\end{figure}

\begin{figure}[htb]
\centering
\includegraphics[width=6.3cm, angle=-90]{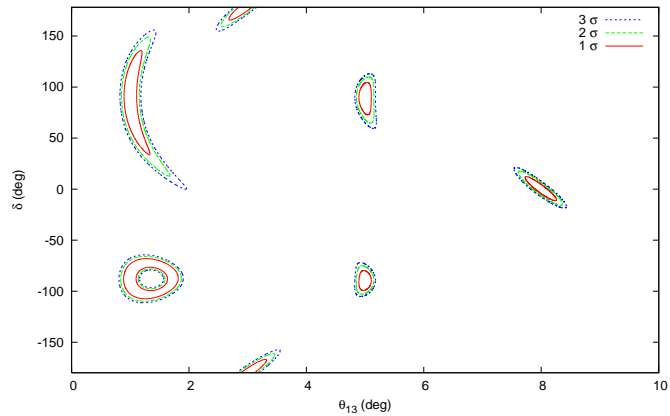}
\caption{\label{fit-EC} The same as Figure 2 for Setup II as explained in the text.}
\end{figure}
For the study of the physics reach associated with such a facility, we combine two different energies for the same $^{150}Dy$ ion in each of two Setups. In all cases we consider $10^{18}$ decaying ions/year, a water Cerenkov Detector with fiducial mass of $440$~Kton and both appearance ($\nu_{\mu}$) and disappearance ($\nu_e$) events. Setup I  is associated with a five year run at $\gamma=90$ (close to the minimum energy to avoid atmospheric neutrino background) plus a five year run at $\gamma=195$ (the maximum energy achievable at present SPS), with a baseline $L=130$~Km from CERN to Frejus. The results for Setup I are going to be compared with those for Setup II, associated with a five year run at $\gamma=195$ plus a five year run at $\gamma=440$ (the maximum achievable at the upgraded SPS with proton energy of $1000$~GeV), with a baseline $L=650$~Km from CERN to Canfranc. As explained in Section 2, the virtues of having at least two energies in a given Setup are that the two oscillation parameters $\theta_{13}$ and $\delta$ can be separated out.

\begin{figure}[htb]
\centering
\includegraphics*[width=7.2cm, angle=-90]{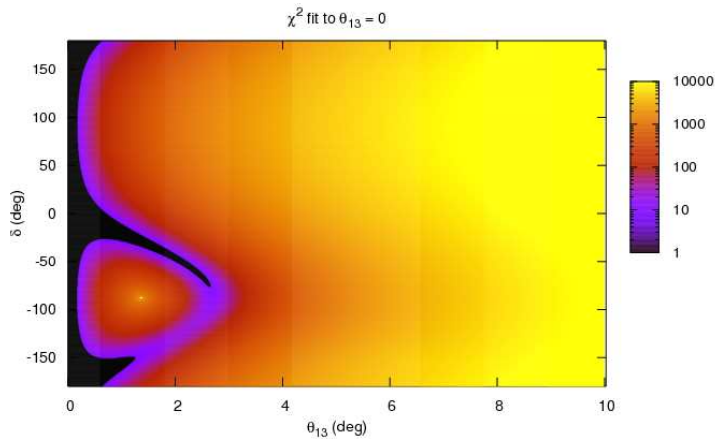}
\caption{Setup II. Sensitivity to  $\theta_{13} \ne 0$ for a two parameter ($\theta_{13}$, $\delta$) fit to the statistical distribution of events.}
\label{sen-EC-theta}
\end{figure}

For the Setup I we generate the statistical distribution of events from  assumed values of $\theta_{13}$ and $\delta$. The corresponding fit with two parameters  is shown in Fig.~\ref{fit-setupI} for selected values of $\theta_{13}$ from $8^o$ to $1^o$ and covering a few values of the CP phase $\delta$. As observed, the principle of an energy dependent measurement (illustrated here with two energies) is working to separate out the two parameters. With this configuration, the precision obtainable for the mixing is much better than that for the CP phase. As seen, even mixings of $1^o$ are still distinguishable. We emphasize that these results are obtained with a two-parameter fit, i.e., assuming that both $(\theta_{13}, \delta)$  are unknown quantities.

At the time of the operation of this proposed Facility, it could happen that the connecting mixing $\theta_{13}$ is already known from the approved experiments for second generation neutrino oscillations, like Double CHOOZ, Daya-Bay, T2K and NOVA. To illustrate the gain obtainable in the sensitivity to discover CP violation from the previous knowledge of $\theta_{13}$, we have reanalyzed the statistical distribution of events with the assumption of $\theta_{13}$ already known in advance. In general, the precision to obtain $\delta$ is then much better than that of Fig.~\ref{fit-setupI} and the corresponding sensitivity for a CP-violation discovery is discussed later.
 
\begin{figure}[htb]
\centering
\includegraphics[width=6cm, angle=-90]{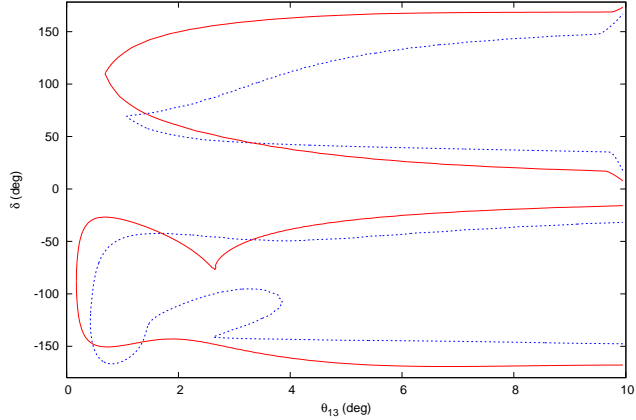}
\caption{\label{fit-EC-Des}CP violation exclusion plot at 99$\%$ CL, if $\theta_{13}$ is still unknown, for the two reference Setups: I (broken blue line) and II (continuous red line).}
\end{figure}
\begin{figure}[htb]
\centering
\includegraphics[width=6cm, angle=-90]{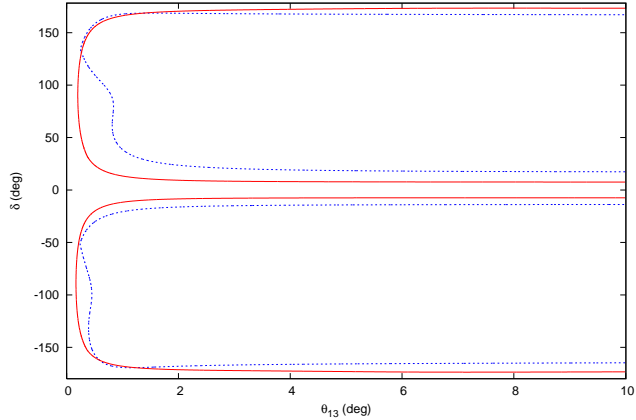}
\caption{\label{fit-EC-Con}CP violation exclusion plot at 99$\%$ CL, if $\theta_{13}$ is previously known, for the two reference Setups: I (broken blue line) and II (continuous red line).}
\end{figure}

In the case of Setup II the longer baseline for $\gamma=195$ leads to a value of $E/L$ well inside the second oscillation (see Fig.~\ref{proba}). In that case the associated strip in the ($\theta_{13}$, $\delta$) plane has a more pronounced curvature, so that the two parameters can be better disantangled. The statistical distribution generated for some assumed values of ($\theta_{13}$, $\delta$) has been fitted and the $\chi^2$ values obtained. The results are given in Fig.~\ref{fit-EC} for a two-parameter fit. Qualitatively, one notices that the precision reachable for the CP phase is better than that in the case of  Setup I. One should emphasize that this improvement in the CP phase determination  has been obtained with the neutrino channel only, using two appropriate different energies. One may discuss in this Setup II the sensitivity to discover $\theta_{13} \ne 0$ by giving the $\chi^2$ fit, for each $\theta_{13}$, to the value $\theta_{13}=0$. This is given  in Fig.~\ref{sen-EC-theta}. Although it is somewhat dependent on the $\delta$-value, we see that values of $\theta_{13}>1^o$ are in general distinguishable for zero.

The corresponding exclusion plots for CP violation in the two Setups are compared when both $\theta_{13}$ and  $\delta$ are unknown. The sensitivity to discover CP violation has been studied by obtaining the $\chi^2$ fit for $\delta=0, 180 ^o$ if the assumed value is $\delta$. For $99\%$~CL, the sensitivities to see CP violation in both Setups are compared in Fig.~\ref{fit-EC-Des}. In both cases, we assume a two-parameter fit, i.e., $\theta_{13}$ previously unknown. For Setup I, a non-vanishing CP violation  becomes significant for $\theta_{13}>4^o$, with values of  the phase $\delta$ around $30^o$ or larger  to be distinguished from zero. For Setup II, the sensitivity to CP violation is better and significant even at $1^o$ in certain cases, depending on the hemisphere for the value of the phase $\delta$.

\begin{figure}[htb]
\centering
\includegraphics*[width=7cm, angle=-90]{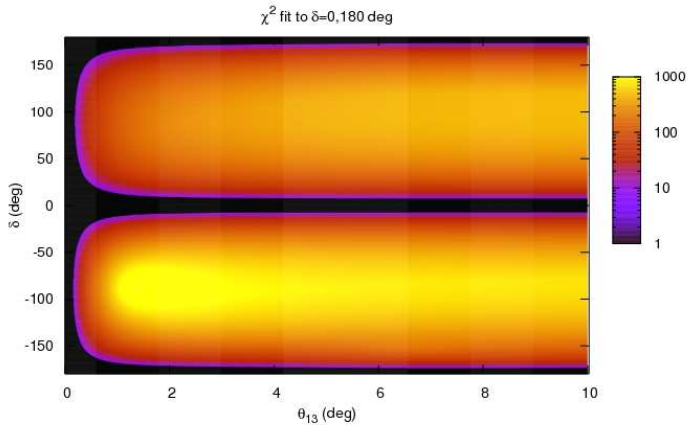}
\caption{Setup II. CPV sensitivity for the statistical distribution of events 
de\-pen\-ding on a single parameter $\delta$, assuming previous information on the value of $\theta_{13}$.}
\label{sen-delta-1par}
\end{figure}

If $\theta_{13}$ is previously known, the corresponding analysis for the sensitivity to discover CP violation is presented in Figure \ref{fit-EC-Con}. In this case, the $\chi^2$ fit is made with the single parameter $\delta$. One may notice that the improvement in this sensitivity is impressive, suggesting that  going step by step in the determination of the neutrino oscillation parameters by means of several generation experiments  is very rewarding. As in Fig.~\ref{fit-EC-Des},  Setup II provides  better sensitivity to the discovery of CP violation than Setup I. In the best case, i.e., $\theta_{13}$ already known  at the time of the proposed experiment with Setup II, we give in Fig.~\ref{sen-delta-1par} the sensitivity to discover  CP violation for different $\chi^2$ to be distinguished from $\delta=0, 180 ^o$. The result is so good that it enters into the regime of a precision experiment.

\vspace{0.5cm}

\section{Conclusions and Outlook}

\noindent The simulations of the physics output for an EC neutrino beam at different energies indicate:

1) The principle of energy dependence to separate out the CP-even and CP-odd contributions to the neutrino oscillation probability works.

2) The upgrade to higher energy in the SPS boost ($E_p=1000$~GeV) helps to have a better sensitivity to CP violation, which is the main objective of the next generation neutrino oscillation experiments, iff accompanied by a longer baseline.

3) The best $E/L$ in order to have a  higher sensitivity to the mixing $\vert U(e3) \vert$ is not the same than that for the CP phase $\delta$. Like the phase-shifts in interference phenomena, the presence of $\delta$ is easier to observe when the energy of the neutrino beams enters  into the region of the second oscillation. The mixing is better seen around the first oscillation maximum, instead.

4) The previous knowledge on the connecting mixing $\theta_{13}$ would greatly improve the sensitivity to CP violation discovery in this facility. This statement is valid in both experimental Setups: I of lower energy, shorter baseline, or II of higher energy, longer baseline.

5) In the best configuration, i.e., with $\theta_{13}$ known in advance and Setup II, the CP-violation sensitivity  is of a few degrees in  $\delta$ for $\theta_{13} \ge 1^o$.

Besides the feasibility studies for the machine, most important for physics is the study of the optimal configuration by combining low energy with high energy neutrino beams, short baseline with long baseline and/or EC monochromatic neutrinos with $^6He$ $\beta^-$ antineutrinos. 

Among the possible systematics associated with the proposed experiments, one should  define a program to determine independently the relevant cross sections of electron and muon neutrinos and antineutrinos with water in the relevant energy region from several hundreds of MeV's to 1 GeV or so.

The result of the synergy of Neutrino Physics with Nuclear Physics (EURISOL) and LHC Physics (SPS upgrade) for the Facility at CERN could be completed with the synergy with Astroparticle Physics for the Detector, which could be  common to neutrino oscillation studies with terrestrial beams, atmospheric neutrinos (sensitive to the neutrino mass hierarchy through matter effects \cite{Bernabeu:1999ct}), Supernova neutrinos and Proton decay. 

The analysis shown in this paper indicates that the proposals discussed here merit $R \& D$ studies in the immediate future for all their ingredients: Facility, Detector and Physics.

\vspace{0.5cm}
\section*{Acknowledgements}
We are indebted to J.~Burguet-Castell, M.~Lindroos and S.~Palomares-Ruiz for interesting discussions. The financial support of this work by the MEC and FEDER Grant FPA 2005-01678 is acknowledged.



\end{document}